# Search for Nearby Ultracool Compact Binaries in Gaia DR2

## D. V. Denisenko[1*], I. Larin[2]

[1] *Sternberg Astronomical Institute, Lomonosov Moscow State University, Russia*
[2] *Education Center on Donskaya, Moscow, Russia*

The information contained in the Gaia Second Data Release (DR2) allows to search for the unusual objects with the pre-selected properties, literally constructing the stars with the desired characteristics. This work describes the idea of method, its implementation and the results of searching for the ultracool compact binaries within 200 parsec distance. The search in the selected range of absolute magnitudes (M=11..13) has yielded the discovery of both expected and unexpected objects. We found two cataclysmic variables with the extreme IR versus UV color index at 126 and 103 parsecs, two eclipsing pre-cataclysmic systems in the period gap and a candidate young stellar object. We analyze the new variable stars using the data from synoptic surveys and compare their properties to the known stars of these types. Applications to the galactic population density of compact binaries are also discussed.

*Key words:* stars: variable, binary, cataclysmic; methods: data mining, multiwavelength, catalogs

## INTRODUCTION

The search described in this work was initiated by the serendipitous discovery of unusual cataclysmic variable Larin 2 in the field centered at NGC 4706 galaxy (Larin et al., 2018). It was found by the 7$^{th}$ grade school student Ivan Larin on the images obtained remotely with iTelescope.Net T31 instrument in Siding Spring. The new variable stood out among the known CVs with the extremely high (*NUV-W1*) color index of 8.4. It was suspected to be a low accretion rate polar similar to MQ Dra (Szkody et al., 2008). However, following the publication of ATel #11401 the new variable Larin 2 has shown an outburst by ~2 magnitudes in ASAS-SN (Shappee et al., 2014; Kochanek et al., 2017) from an average level of 17.5$^m$ to 15.7$^m$ (Figure 1). Such outbursts were never observed in MQ Dra over nine years of Catalina Realtime Transient Survey (Drake et al., 2009).

It became obvious that Larin 2 is an extreme object with the exceptional properties that has escaped the detection by synoptic surveys due to the low outburst amplitude and the lack of periodic variations. Most transient search projects are programmed to detect the objects which show brightening (or fading) by at least 2.0, 2.2 or even 2.5 magnitudes. As a result, low amplitude dwarf novae remain undetected among other non-periodic variable stars of potential astrophysical interest.

Now that we know what to search for, we decided to look for other objects with the properties similar to Larin 2 in Gaia DR2. It would be very unlikely that the object that has shown up in the randomly selected search field less than 1 square degree (FOV of iTelescope T31 is 56'×56') is the only one of the kind in the whole sky. According to Gaia DR2, Larin 2 has an absolute magnitude M=12.2 (*Gmag*=17.70 at 126 pc distance) and color index *BP-RP*=2.7, well beyond the locus of CVs in Fig. 6 of Eyer et al., 2018.

---

* E-mail: d.v.denisenko@gmail.com

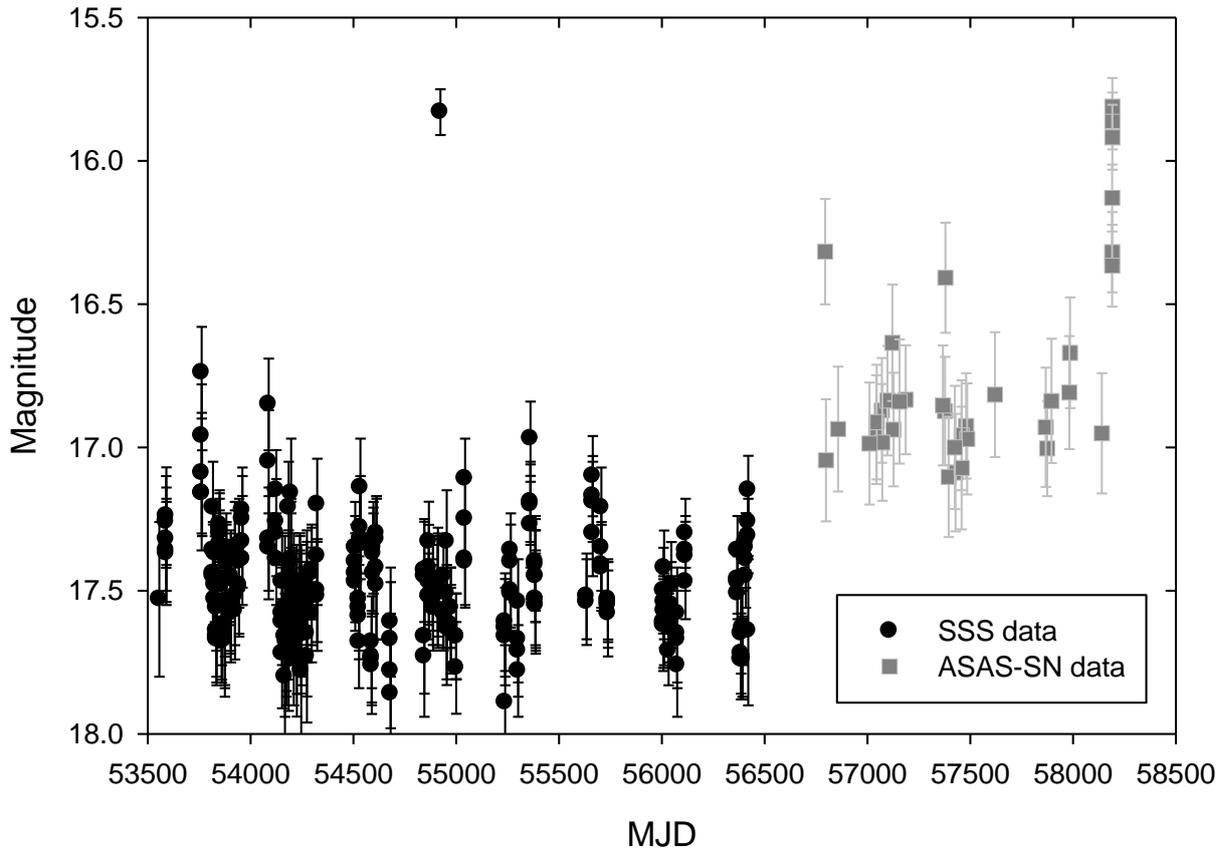

**Figure 1.** Light curve of Larin 2 from Siding Spring Survey (black circles) and ASAS-SN data (gray squares). ASAS-SN upper limits for non-detections are not shown.

## SEARCH METHOD

The search for unusual variable objects was performed in three steps:

1) TAP Vizier query to Gaia DR2 (I/345) limited to red stars with absolute magnitude M from 11 to 13 within 200 pc distance (parallax of 5 mas or larger)
2) Vizier query of the list from Step 1 to GALEX catalogue (II/312)
3) Checking the objects from Step 2 with *NUV* brighter than 22 for variability in the publicly available CRTS and ASAS-SN databases

The following query was submitted to http://tapvizier.u-strasbg.fr/adql/ for the 30°×30° box-shaped fields separated by 30 degrees (30-degree step in R.A. at -30°, 0° and +30° declination and 60-degree step in R.A. at -60° and +60° declination):

```
-- output format : csv
SELECT "I/345/gaia2".ra, "I/345/gaia2".dec, "I/345/gaia2".parallax,
"I/345/gaia2".pmra, "I/345/gaia2".phot_g_mean_mag, "I/345/gaia2".bp_rp,
"I/345/gaia2".pmdec
FROM "I/345/gaia2"
WHERE 1=CONTAINS(POINT('ICRS',"I/345/gaia2".ra,"I/345/gaia2".dec), BOX('ICRS',
210.000, 00.000, 30., 30.))
AND (bp_rp<3) AND (bp_rp>2) AND (parallax>5) AND (phot_g_mean_mag-
5*log(1000/parallax)/log(10)+5)>11 AND (phot_g_mean_mag-
5*log(1000/parallax)/log(10)+5)<13
```

Here `210.000, 00.000` corresponds to the field centered at R.A. = 14$^h$ and Decl. = 0°. The set of 50 fields is covering the whole celestial sphere (twelve at -30°, 0° and +30°, six at -60° and +60° plus two polar caps centered at -90° and +90° with a radius of 15 degrees).

While Gaia DR2 (Brown et al., 2018) covers the entire sky, this is not the case for GALEX and CRTS. Both surveys are avoiding dense Milky Way fields. Catalina Realtime Transient Survey (Drake et al., 2009) is limited in declination by +70° at North (Catalina Sky Survey, CSS, and Mount Lemmon Survey, MLS) and by -75° at South (Siding Spring Survey, SSS). GALEX All-Sky Imaging Survey (Bianchi et al., 2011) has some gaps irregularly distributed over the sky. That puts the inevitable limitations on our search process. Obviously, we are missing the potentially interesting objects within 10-15° of the Galactic plane.

Each 30°×30° field had on average about 5000 stars in Gaia DR2 satisfying the criteria. Only about one per cent of them had UV detection in GALEX. The remaining several hundred objects were queried using CRTS photometry database. CRTS light curves were inspected visually, and the objects showing remarkable variations above the mean error corridor were analyzed further. The search for the periodic variability was then carried out in WinEffect software by Dr. V. P. Goranskij.

## SEARCH RESULTS

In addition to Larin 2, we have found four new variable stars with the specified properties. The list of objects found during our work is given in Table 1. Stars are listed in the chronological order of discovery. Finder charts for DDE variables are given at the website http://scan.sai.msu.ru/~denis/VarDDE.html.

| Star | Gaia Position | Gaia Plx | Gmag | BP-RP | FUV | NUV | NUV-W1 |
|---|---|---|---|---|---|---|---|
| Larin 2 | 12 48 50.81 -41 26 54.4 | 7.95±0.13 | 17.70 | 2.72 | 21.31 | 21.42 | 8.4 |
| DDE 157 | 16 42 51.55 +01 35 54.8 | 10.88±0.09 | 16.63 | 2.35 | N/A | 20.18 | 7.7 |
| DDE 158 | 14 09 34.64 -38 46 10.5 | 9.71±0.10 | 16.09 | 2.62 | 20.46 | 20.18 | 8.9 |
| DDE 159 | 02 52 59.75 +09 44 12.2 | 11.24±0.10 | 16.51 | 2.95 | N/A | 21.25 | 9.5 |
| DDE 160 | 00 08 12.06 +18 53 45.7 | 5.27±0.17 | 17.62 | 2.36 | N/A | 20.99 | 7.3 |

**Table 1.** Data for the newly discovered variable stars. Coordinates (Epoch 2015.5, Equinox J2000.0), parallaxes, *G* magnitudes and *BP-RP* color indices are from Gaia DR2, *FUV* and *NUV* from GALEX and *W1* from WISE.

Table 2 gives the additional information about new objects: their CRTS IDs, magnitude ranges, variability types, periods (when available) and constellations.

| Star | CRTS designation | Mag range | Type | Period, d | Constellation |
|---|---|---|---|---|---|
| Larin 2 | SSS J124850.8-412654 | 15.7-17.8 | CV (DQ/UG?) | Unknown | Centaurus |
| DDE 157 | CSS J164251.5+013554 | 16.3-16.9 | EA+WD | 0.0962907 | Ophiuchus |
| DDE 158 | SSS J140934.6-384610 | 15.4-16.2 | CV (UG?) | Unknown | Centaurus |
| DDE 159 | CSS J025259.7+094412 | 15.8-17.0 | YSO | Irregular | Cetus |
| DDE 160 | CSS J000812.1+185345 | 17.5-17.9 | EA | 0.126729 | Pegasus |

**Table 2.** Variability details of the newly discovered stars. Magnitude ranges are from Catalina Realtime Transient Survey (unfiltered with the *V* zero point).

# NOTES ON INDIVIDUAL OBJECTS

**Larin 2.** As mentioned above, this object is nearly a twin to the low accretion rate polar (LARP) MQ Dra = J1553.5+5516. Table 3 lists the properties of two stars in comparison.

| Object | FUV | NUV | B2 | R2 | J | H | K | W1 | W2 | W3 | W4 |
|---|---|---|---|---|---|---|---|---|---|---|---|
| MQ Dra | 21.30 | 21.04 | 18.69 | 17.26 | 14.59 | 14.04 | 13.76 | 13.39 | 12.96 | 11.44 | 9.63 |
| Larin 2 | 21.31 | 21.42 | 18.80 | 17.85 | 14.23 | 13.59 | 13.25 | 13.01 | 12.52 | 11.49 | 8.93 |

**Table 3.** Ultraviolet (GALEX *FUV* and *NUV*), visible (USNO-B1.0 *B* and *R*), near infrared (2MASS *JHK*) and infrared (WISE *W1-W4*) magnitudes of MQ Dra and Larin 2.

Moreover, their distances are also quite similar (186 pc for MQ Dra and 126 pc for Larin 2), implying the absolute magnitudes differing by no more than $1.0^m$ in optical and $0.5^m$ in infrared. The 4-hour long time series with 0.5-m Chilescope on Mar. 10 (Larin et al., 2018) has shown the orbital variation by at least $0.5^m$, from 16.0 to 16.5, with a possible indication of white dwarf spin period about 0.0362 d (52.1 min). MQ Dra itself varies between 17.1 and 17.6 unfiltered magnitude in CRTS data. Thus, the difference between the absolute magnitudes of MQ Dra and Larin 2 is well within their variability amplitudes. Together with the nearly identical colors this implies similar physical properties of two systems. MQ Dra has an orbital period of P=0.182970 d (4.4 hour) and the secondary component of M5 spectral type. Time resolved spectroscopy of Larin 2 is required to determine the orbital period and mass ratio of components. The monitoring for future outbursts is also strongly encouraged.

**DDE 157.** This is an eclipsing WDMS (white dwarf + red dwarf) variable with the period of 2.311 hr whose orbital light curve is not changing with time during nine years of CRTS observations. Thus, there is currently no interaction between the components. We consider it to be a pre-cataclysmic variable crossing the period gap. See Zorotovic et al., 2016 for the review of detached CVs. DDE 157 is the first eclipsing WDMS system in the period range from 2.2 to 3.0 hours. It is obvious from the large *NUV-W1* color that the white dwarf is cool. Comparing its parameters to those of other eclipsing WDMS systems, one can estimate the temperature of the white dwarf. *NUV-W1* value of DDE 157 (7.7) is close to that of WD 0300+007 (7.8) which has $T_{WD}<8000$ K, according to Zorotovic et al.

The phased light curve of DDE 157 is presented at Figure 2. The difference between DDE 157 and majority of other WDMS binaries is the large amplitude of sine wave outside the eclipses indicating the significant elongation of red dwarf component. It is generally accepted that the accretion resumes as the binary system approaches the lower boundary of the period gap. DDE 157 can serve as the laboratory for studying the evolution of pre-cataclysmic variables. Its eclipses are nearly head-on, allowing to measure directly the physical parameters of both components. High-speed photometry and time-resolved spectroscopy are encouraged. The heliocentric eclipse ephemeris is as follows:

$$HJD = 2456559.598 + 0.0962907 \times E$$

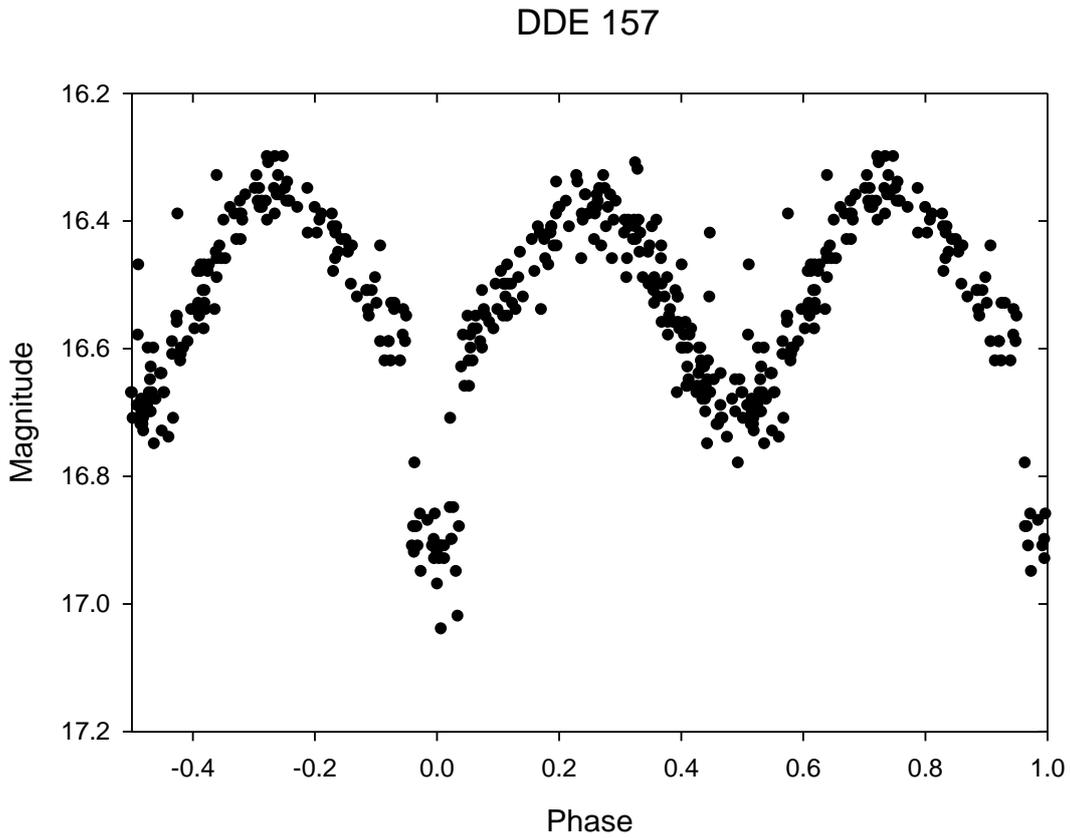

**Figure 2.** Light curve of DDE 157 from Catalina Sky Survey folded with the best orbital period P=0.0962907 d (2.311 hr).

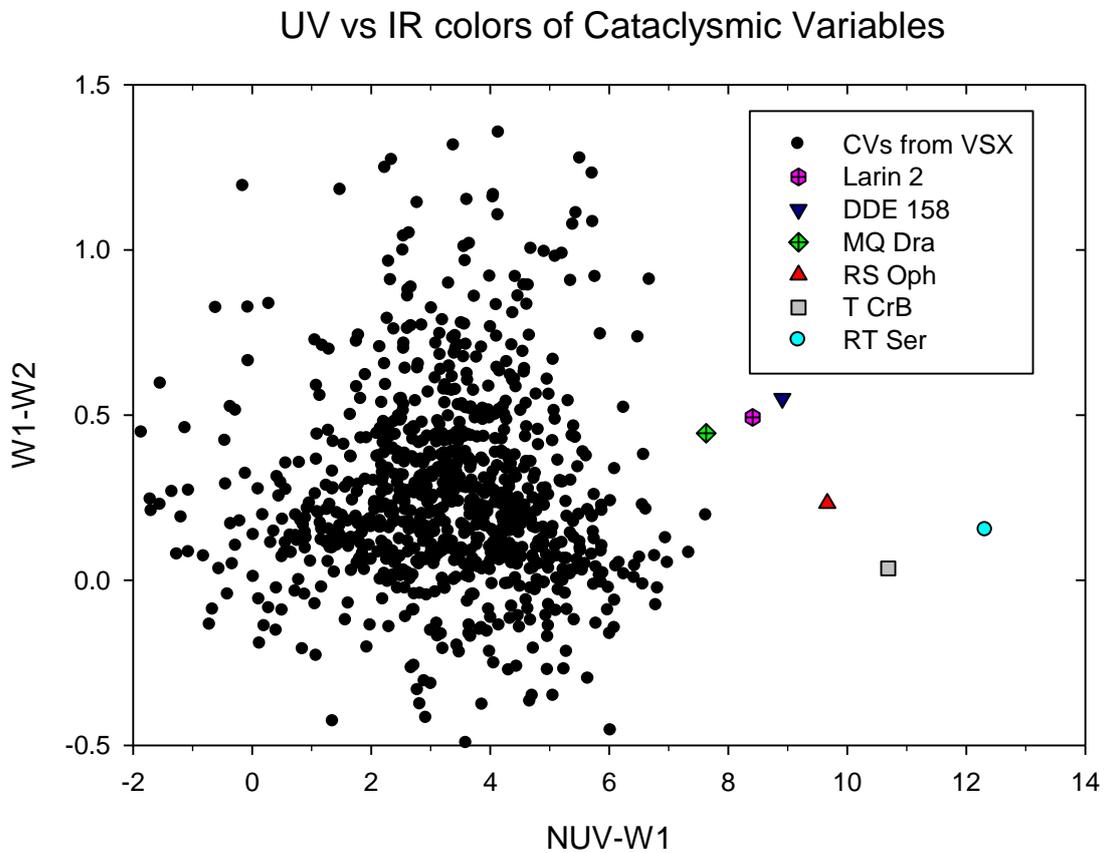

**Figure 3.** Location of Larin 2 and DDE 158 on the UV vs IR color diagram of cataclysmic variables with known *NUV*, *W1* and *W2* magnitudes in GALEX and WISE catalogs.

**DDE 158.** This variable star is probably the most remarkable one among the new objects reported here. Its *NUV-W1* value of 8.9 is record high for all cataclysmic variable stars, except for recurrent novae. Figure 3 shows the location of Larin 2 and DDE 158 on the IR vs UV color diagram for all CVs from AAVSO VSX with the measured *NUV*, *W1* and *W2* values in GALEX and WISE catalogues, respectively. It is obvious that MQ Dra, Larin 2 and DDE 158 stand out from the main locus of CVs. Only recurrent novae which are principally different objects compared to classical CVs have larger UV-IR colors.

Light curve of DDE 158 from Siding Spring Survey is presented at Figure 4. The variable is showing rather frequent outbursts with an amplitude of less than $1^m$, from 16.2 to 15.4.

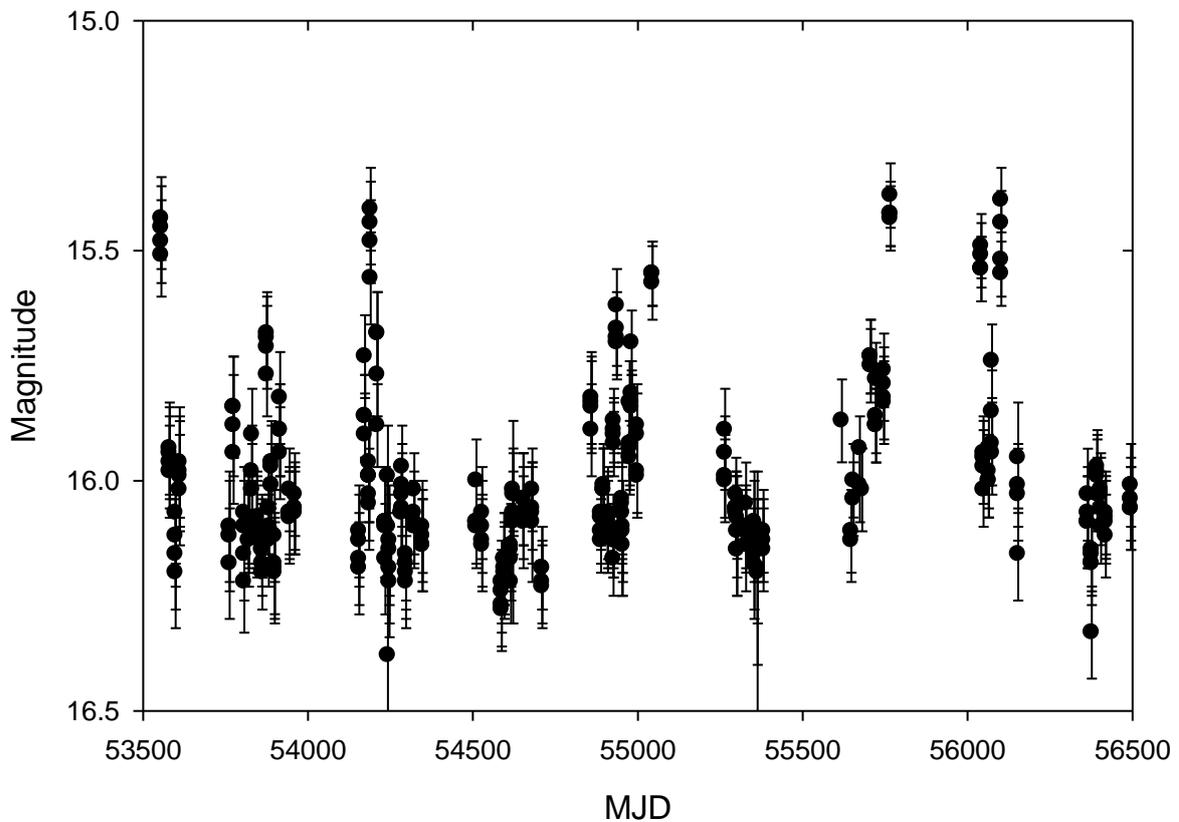

**Figure 4.** Light curve of DDE 158 = SSS J140934.6-384610 from Siding Spring Survey

*FUV-NUV* color of 0.28 confirms the presence of a white dwarf in the system. Light curve of DDE 158 is typical for dwarf novae, except for the outburst amplitude. Most of U Gem type CVs have outbursts by 2-6 magnitudes and reach the absolute magnitude of 4.5-5 at the peak. DDE 158 has unfiltered M=11.1 at quiescence and M=10.3 in outburst.

We note there is a faint X-ray source 1RXS J140927.0-384650 formally 98" away from DDE 158 with the flux of 0.037±0.015 cts/s and hardness ratios HR1=1.00±0.31, HR2=0.07±0.34 (Voges et al., 2000). However, it is likely associated to the quasar within ROSAT error circle, 6" from the nominal X-ray position, which is brighter than DDE 158 in UV by $1.5^m$ (*FUV*=19.21, *NUV*=18.51).

With Gaia distance of 103 parsec DDE 158 is among Top 20 nearest cataclysmic variables to us, as shown in Table 4. Yet it is the closest CV with unknown orbital period. There is a possible periodic signal in SSS data at P=0.144693 d. The further study of this intriguing object is highly recommended.

| Star | Distance, pc | Mag range | Type | Period, d | Constellation |
|---|---|---|---|---|---|
| WZ Sge | 45 | 7 - 15.5 | UGWZ+E | 0.056688 | |
| VW Hyi | 54 | 8.4-14.4 | UGSU | 0.074271 | |
| EX Hya | 57 | 10.0-14.3 | UG/DQ+E | 0.068234 | |
| V455 And | 75 | 8.5-16.5 | UGWZ | 0.056310 | |
| ASASSN-14dx | 81 | 13.9-17.0 | UGSU+E | 0.057506 | Cetus |
| NSV 25966 | 81 | 16.0-16.4 | NL | 0.07134 | Cepheus |
| AM Her | 88 | 12.3-15.7 | AM+E | 0.128927 | |
| AE Aqr | 91 | 10.2-12.1 | DQ+ELL | 0.411656 | |
| IX Vel | 91 | 9.1-10.0 | NL | 0.193927 | |
| OY Car | 91 | 11.2-16.2 | UGSU+E | 0.063121 | |
| U Gem | 93 | 8.2-14.9 | UGSS+E | 0.176906 | |
| BW Scl | 94 | 8.9-17.2 | UGWZ | 0.054323 | |
| WX LMi | 99 | 15.8-17.0 | AM | 0.115924 | |
| V627 Peg | 100 | 8.8-16.3 | UGWZ | 0.054523 | |
| AR UMa | 101 | 14.2-15.9 | AM | 0.080501 | |
| DDE 158 | 103 | 15.4-16.2 | UG: | Unknown | Centaurus |
| Gaia14abg | 106 | 17.5-17.9 | CV: | 0.065327 | Hercules |
| BPM 18764 | 107 | 15.4-15.8 | NL | Unknown | Carina |
| V2051 Oph | 112 | 11.6-17.5 | UGSU | 0.062428 | |
| V834 Cen | 112 | 14.2-17.0 | AM | 0.070498 | |

**Table 4.** Twenty closest cataclysmic variables. Distances are from Gaia DR2, magnitude ranges and types are from AAVSO VSX (Watson et al., 2006; B/vsx).

**DDE 159.** This object is showing fading episodes by $1.2^m$, but without a single definite period. Lafler-Kinman method detects several periodic signals at 0.184387, 0.217353, 0.269720 d and their daily aliases. However, the phased light curves for all of them are unsatisfactory. This is likely a young stellar object at 89 pc. The light curve of DDE 159 from Catalina Sky Survey is presented at Figure 5.

**DDE 160.** This is another eclipsing white dwarf + red dwarf variable with the period of 3.04 hr. The phased light curve of DDE 160 is presented at Figure 6. It differs remarkably from that of DDE 157. There is almost no sinusoidal variation. There is an increased number of eclipsing WDMS systems at the upper boundary of the CV period gap. Four out of 22 such systems in Zorotovic et al., 2016 are falling in the bin from 3.0 to 3.2 hours. Parsons et al., 2015 are listing 71 eclipsing WDMS systems, 11 of them having periods less than 0.13 d. Drake et al., 2014 have published the spectra of 19 ultra-short period binaries from Catalina Surveys, but all of them are in the interval from 0.139 to 0.212 d. Our finding shows that even shorter period binaries can be found in CRTS data. The spectroscopic study of DDE 160 is encouraged. Light elements of eclipses:

$$HJD = 2454833.614 + 0.1267294 \times E$$

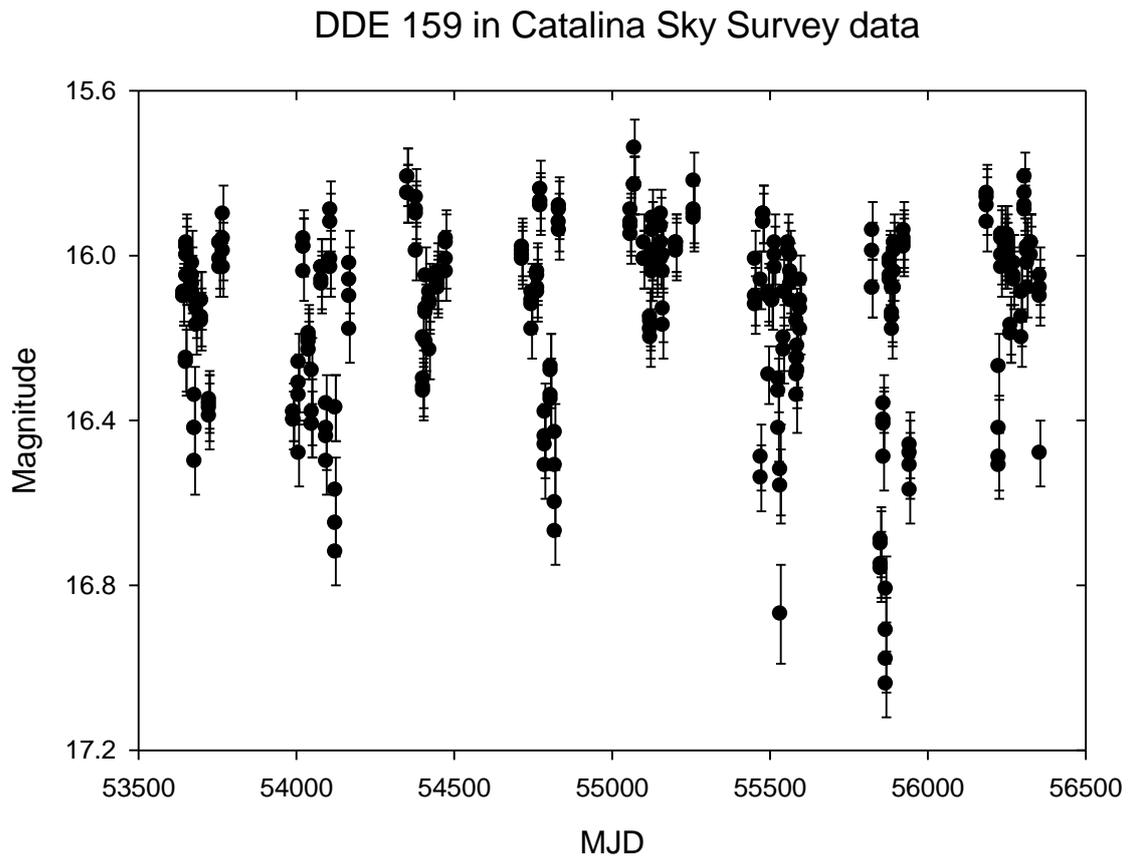

**Figure 5.** Light curve of DDE 159 = CSS J025259.7+094412 from Catalina Sky Survey

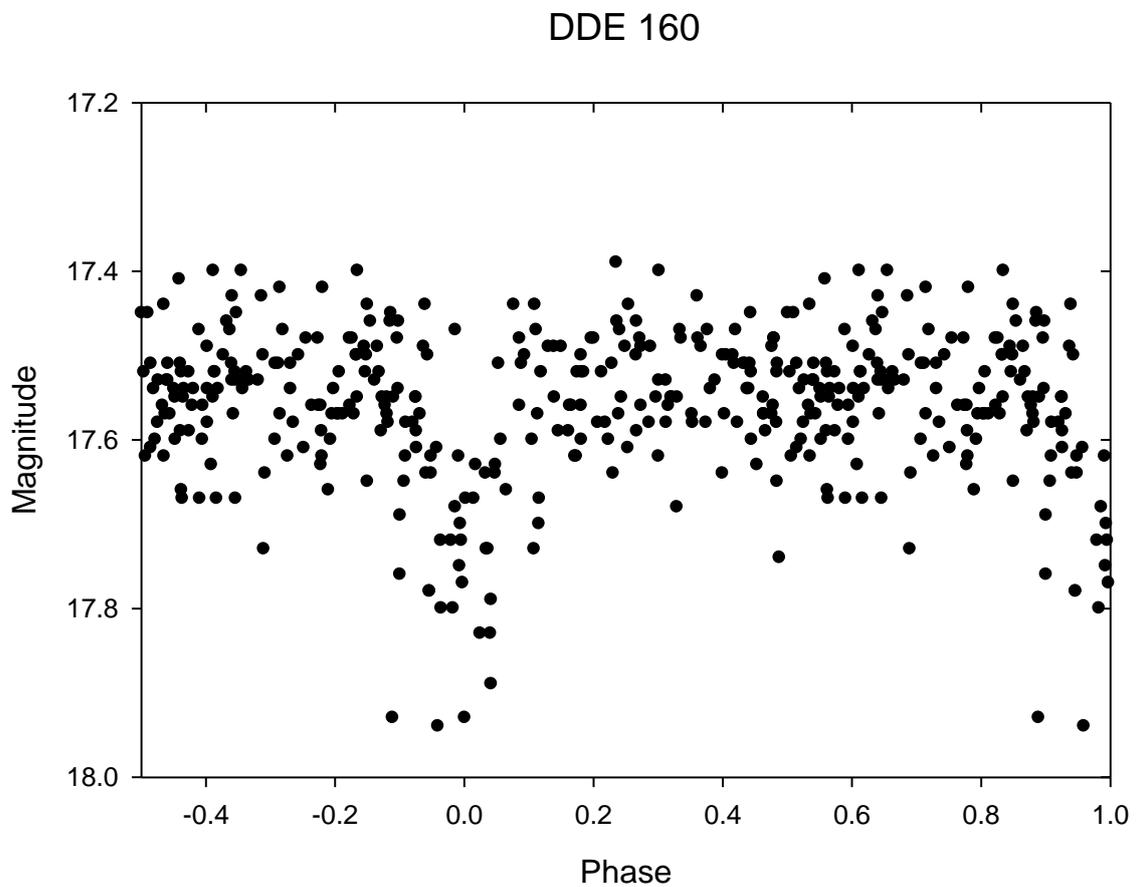

**Figure 6.** Light curve of DDE 160 from Catalina Sky Survey folded with the best orbital period P=0.1267294 d (3.0415 hr).

DISCUSSION

Gaia Second Data Release has provided us with the invaluable data on the distances and magnitudes of about a billion stars in our Galaxy. Using this data, we can improve our knowledge of various galactic populations and even find the previously unknown types of objects. The search presented in this work was aimed at the nearby objects that are blue in ultraviolet and extremely red in optical and infrared bands. By constructing the proper query to catalogs we have found two binary systems with unprecedented *NUV-W1* colors for cataclysmic variables. One of our discoveries is located at 103 pc distance, ranking 16$^{th}$ in the list of nearest CVs. Two objects (Larin 2 and DDE 158) are showing low amplitude outbursts (1.8 and 0.8$^m$, correspondingly) making them nearly impossible to discover by the present ground based synoptic surveys. Besides the small variability, both objects are intrinsically faint (M~12) which means their analogs can only be found within a few hundred parsecs from the Sun. Since our search did not cover the dense parts of the Milky Way, there must be other similar systems escaping the detection. Gaia is opening new horizons for identifying the objects like these and for refining the space density of compact binary systems of various kinds (detached, semi-detached WDMS binaries and interacting cataclysmic variables).

Given the low number of the newly discovered variable stars it is hard to quantitatively estimate the contribution of the low luminosity objects to the galactic population of compact binary systems. Since all our findings are located at the galactic latitudes larger than ±20, there should be at least as many objects within 20 degrees from the Galactic plane. Being extremely cautious, one can say that the population of cataclysmic binaries is underestimated by 10-20 per cent just due to the lack of discovered low accretion rate systems.

Long-term light curves of Larin 2 and DDE 158 can be compared to that of low-amplitude dwarf nova V364 Lib (Kimura et al., 2018). However, V364 Lib is a completely different object with an absolute magnitude M=2.65 (Gaia distance 565 pc, *Gmag*=11.41). The detailed spectroscopic and photometric investigation of these two variables is required to uncover their true nature. They are probably the members of a new class of outbursting cataclysmic variables with the extremely low accretion rate (*micronovae* or *nanonovae*). Search for other objects of this type can be performed in the data of synoptic sky surveys.

APPENDIX: STARS WHICH ARE NOT CVs

During the preparation of Figure 3 and Table 4 we have found several objects which are listed in AAVSO VSX as being CVs or possible CVs but are not in fact cataclysmic variables. We report them here with the correct ID and classification, when possible.

2MASS J20310630-4914562 (Plavchan et al., 2008): eclipsing variable of EW type with P= 0.274044 d (Drake et al., 2017).

ASASSN-16bd (Shappee et al., 2014): QSO, variable radio source, z=0.73151 in 6dF galaxy survey (Jones et al., 2009).

CSS 131106:032129+180827 (Drake et al., 2009): 78.5 pc, *Gmag*=19.22, M=14.7 – flaring red dwarf (UV Ceti type). CSS light curve shows one flare on 2013-11-06 peaking at $16.81^m$, fading to $17.57^m$ within an hour and further to $18.09^m$ in two hours.

MASTER OT J100950.32+471815.8 (Balanutsa et al., 2012): 142 pc, *Gmag*=18.26, M=12.5 – red dwarf. Red object (*r*=19.31, *i*=17.52) detected on the unfiltered image at $18.0^m$, while the reference image was in R filter with limiting magnitude 18.8.

PS15ber (Wright et al., 2015): 33.0 pc, *Gmag*=17.55, M=15.0 – high proper motion star (2.3"/yr) not present on PanSTARRS reference images.

ACKNOWLEDGMENTS

This research has made use of the VizieR catalogue access tool, CDS, Strasbourg, France (Ochsenbein et al., 2000) and its mirror at CfA/Harvard, USA. This work has made use of data from the European Space Agency (ESA) mission Gaia (https://www.cosmos.esa.int/gaia), processed by the Gaia Data Processing and Analysis Consortium (DPAC, https://www.cosmos.esa.int/web/gaia/dpac/consortium). Authors thank Dr. V. P. Goranskij for making WinEffect software freely available and Dr. K. V. Sokolovsky for his online Period search service at http://scan.sai.msu.ru/lk/.